\documentclass[aip,apl,numerical,preprint]{revtex4-1}

\usepackage{graphicx}%
\usepackage{bm}%
\usepackage{amssymb}
\usepackage[amssymb]{SIunits}
\usepackage[version=3]{mhchem}

\newcommand{\CNRSAddress}{CEA/CNRS/UJF joint team ``Nanophysics and Semiconductors'', Institut N\'eel-CNRS, BP 166, 25 rue des Martyrs, 38042 Grenoble Cedex 9, France}
\newcommand{\CEAddress}{CEA/CNRS/UJF joint team ``Nanophysics and Semiconductors'', CEA/INAC/SP2M, 17 rue des Martyrs, 38054 Grenoble cedex 9, France}

\begin{document}

\title{Surface effects in a semiconductor photonic nanowire and spectral stability of an embedded single quantum dot}

\author{\surname{Inah} Yeo}
\affiliation{\CNRSAddress}
\author{\surname{Nitin S.} Malik}
\affiliation{\CEAddress}
\author{\surname{Mathieu} Munsch}
\affiliation{\CEAddress}
\author{\surname{Emmanuel} Dupuy}
\affiliation{\CEAddress}
\author{\surname{Jo\"el} Bleuse}
\affiliation{\CEAddress}
\author{\surname{Yann-Michel} Niquet}
\affiliation{Laboratoire de Simulation Atomistique, CEA/INAC/SP2M, 17 rue des Martyrs, 38054 Grenoble cedex 9, France}
\author{\surname{Jean-Michel} G\'erard}
\affiliation{\CEAddress}
\author{\surname{Julien} Claudon}
\email[Electronic mail: ]{julien.claudon@cea.fr}
\affiliation{\CEAddress}
\author{\surname{\'Edouard} Wagner}
\affiliation{\CNRSAddress}
\author{\surname{Signe} Seidelin}
\affiliation{\CNRSAddress}
\author{\surname{Alexia} Auff\`eves}
\affiliation{\CNRSAddress}
\author{\surname{Jean-Philippe} Poizat}
\affiliation{\CNRSAddress}
\author{\surname{Gilles} Nogues}
\email[Electronic mail: ]{gilles.nogues@grenoble.cnrs.fr}
\affiliation{\CNRSAddress}

\date{\today{}}

\begin{abstract}
We evidence the influence of surface effects for \ce{InAs} quantum dots embedded
into \ce{GaAs} photonic nanowires used as efficient single photon sources. We
observe a continuous temporal drift of the emission energy that is an obstacle
to resonant quantum optics experiments at the single photon level. We attribute
the drift to the sticking of oxygen molecules onto the wire, which modifies the
surface charge  and hence the electric field seen by the quantum dot. The
influence of temperature and excitation laser power on this phenomenon is
studied. Most importantly, we demonstrate a proper treatment of the nanowire
surface to suppress the drift. 
\end{abstract}


\maketitle

Efficient nanophotonic devices like single photon sources require to funnel a
large fraction $\beta$ of the spontaneous emission (SE) of a single emitter into
a single optical mode. This situation offers an ideal
platform for quantum optics and quantum information processing
experiments~\cite{GIANTNONLINEARITY_07,
EnglundVuckovic_Controllingcavityreflectivity_07}. In this context, photonic
semiconducting nanowires (PW) embedding a single quantum dot (QD) have emerged
as appealing systems~\cite{FriedlerG'erard_Solid-statesinglephoton_09,
ClaudonGerard_highlyefficientsingle-photon_10,
BabinecLoncar_diamondnanowiresingle-photon_10}. These monomode waveguides made
of a high refractive index material, offer a tight lateral confinement of the
guided mode while simultaneously screening all the other transverse modes. Hence
they ensure an efficient SE control over a broad wavelength range
($\beta\ge$90\% over a bandwidth exceeding \unit{100}{\nano\meter} at $\lambda=
$\unit{950}{\nano\meter})\cite{FriedlerG'erard_Solid-statesinglephoton_09,
NANOWIRESECONTROLBLEUSE11}. Moreover the far-field outcoupling efficiency of
the guided mode can reach unity by proper engineering of the wire
ends\cite{FriedlerG'erard_Solid-statesinglephoton_09}. Following this strategy,
an on-demand single-photon source with a record-high brightness was 
demonstrated\cite{ClaudonGerard_highlyefficientsingle-photon_10}.

However, miniaturization of photonic devices enhances surface effects, inducing
for example non radiative surface recombinations which affect the emission
properties of semiconducting nanowires\cite{DEMICHEL_SURFACEGAAS_10,
SURFACEEFFECTDAN11}. In the case of PWs, single-mode operation imposes a wire
diameter $d\lesssim\lambda/n$, where $n$ is the index of refraction of the
material. For \ce{GaAs} PWs the QD is thus located at distances not larger than
\unit{100}{\nano\meter} from sidewalls. It was shown in
Ref.~\citenum{NANOWIRESECONTROLBLEUSE11} that the almost perfect QD radiative
yield is preserved in PWs with $d$ as small as \unit{200}{\nano\meter}. 
Photon correlation experiments showed no bunching in the 1--\unit{100}{\nano\second} temporal range\cite{ClaudonGerard_highlyefficientsingle-photon_10}, a strong evidence of absence of blinking at this timescale.
Nevertheless, the spectral stability of the PWs, crucial for resonant quantum
optics experiments, has not been investigated so far. In this letter, we perform
high resolution spectroscopy and show that the excitonic emission line of the QD
undergoes a continuous energy drift. We discuss the possible origin of the
phenomenon, which we attribute to oxygen adsorption on the wire sidewalls, and
demonstrate a way to circumvent this problem with a proper surface treatment.

\begin{figure}
\centering
{\includegraphics[height=5.7cm,keepaspectratio=true]{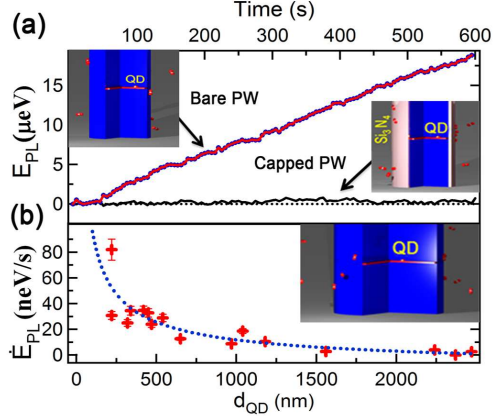}}
 \caption{(Color online) (a) $E_{\rm PL}$ versus time for a QD in a bare PW
(blue line) and  in a capped PW (black line). (b) $\dot{E}_{\rm PL}$ for QDs in
bare PWs versus wire diameter}
 \label{fig:drift}
\end{figure}

The device fabrication starts from a planar structure grown by molecular beam
epitaxy on a \ce{GaAs} wafer. A single layer of \ce{InAs} self-assembled QDs
(areal density \unit{300}{\micro\meter\rpsquared}) is located in a GaAs matrix
(residual positive doping $p=$\unit{\power{10}{16}}{\centi\meter\rpcubed}). The
ensemble luminescence peaks at \unit{920}{\nano\meter} (\unit{50}{\nano\meter}
inhomogeneous broadening). PWs are defined with a top-down approach, using
e-beam lithography and dry plasma etching (\ce{Ar-SiCl4} plasma).  A careful
control of the etching leads to a conical wire geometry in order to control
the optical mode transverse profile along the wire. A series of PWs with top
diameters varying by a fine \unit{10}{\nano\meter} step were fabricated.  The
actual wire diameter at the dot location $d_{\rm QD}$ is measured with a
$\sim$\unit{10}{\nano\meter} accuracy by electron microscopy. A wire with a
diameter $d_{\rm QD}=$\unit{200}{\nano\meter} typically contains 10 randomly
located QDs. As surface passivation helps in reducing surface
effects~\cite{SURFACEEFFECTDAN11}, we start from the previous device to
fabricate surface capped-PWs. The oxide layer is removed by wet chemical
etching. \ce{GaAs} dangling bonds at the surface are saturated by \ce{(NH4)2S}
in order to reduce the density of surface traps. Finally a
\unit{10}{\nano\meter}-thick layer of \ce{Si3N4} is deposited over the surface.

The sample is mounted on a cold finger cryostat with optical access. QDs are
excited through a 0.75 NA microscope objective. Standard cw diode or Ti:Sa
laser excite the QDs. Bandgap (BG) excitation at \unit{1.52}{\electronvolt}
creates electron-hole pairs in the bulk \ce{GaAs} matrix. Wetting layer (WL)
excitation (\unit{1.49}{\electronvolt}) only excites carriers in the continuum
absorption band of the \ce{InAs} monolayer close to the QDs. Finally
quasi-resonant p-shell excitation (\unit{1.42}{\electronvolt}) excites discrete
QD states. The light emitted by the recombination of the QD lowest energy states
is recollected by the same lens and sent to a \unit{1.5}{\meter} focal length
spectrometer (\unit{12}{\micro\electronvolt} spectral resolution at
\unit{1.38}{\electronvolt}). A CCD camera is placed at the output of the
spectrometer. Its pixel size (\unit{20}{\micro\meter}) corresponds to an energy
step comparable to the resolution. Successive QD photoluminescence (PL) spectra
are recorded and fitted with a Lorentzian function. Considering the signal to
noise ratio for the data, we estimate the statistical uncertainty on the peak
position $E_{\rm PL}$  to be \unit{0.1}{\micro\electronvolt}. 

Figure \ref{fig:drift}(a) presents the PL peak position $E_{\rm PL}$ versus time
for a single QD in a $d_{\rm QD}=$\unit{370}{\nano\meter} bare PW at
\unit{3.5}{\kelvin}. The quantum dot is excited by pumping the WL at a power
$P/P_{\rm sat}=0.2$ ($P_{\rm sat}$ is the saturation power of the QD). A regular
drift $\dot{E}_{\rm PL}$ of the emission energy towards higher frequencies is
observed (blue drift). A linear fit of the data of Fig.~\ref{fig:drift}(a) gives
$\dot{E}_{\rm PL}\simeq$\unit{30}{\nano\electronvolt\per\second}. At
\unit{3.5}{\kelvin}, $\dot{E}_{\rm PL}$ remains constant over 8 hours.  
Over this period of time, PL intensity does not change significantly, its linewidth remains limited by the resolution of the spectrometer and lifetime measurements show a small decrease from 1.65 to \unit{1.58}{\nano\second}.
Warming up to room temperature and pumping on the cryostat resets the emission energy to
its initial lower value. A systematic study of many PWs with different
diameters [Fig.~\ref{fig:drift}(b)] reveals that all QDs are affected by a blue
drift whose amplitude decreases with increasing diameter. Finally the same
experiment with QDs in  capped PWs shows no drift (Fig.~\ref{fig:drift}(a),
$\dot{E}_{\rm PL}\le$\unit{0.7}{\nano\electronvolt\per\second}).

Previous observations clearly point towards surface effects. Dry etching of PWs
during their fabrication creates a large density of surface
traps~\cite{DEMICHEL_SURFACEGAAS_10}
$n_s\gtrsim$\unit{\power{10}{12}}{\centi\meter\rpsquared}. For \ce{GaAs}, the
energy of those traps is in the middle of the gap ($\Phi \equiv E_{\rm
trap}-E_{\rm VB}=$\unit{0.7}{\electronvolt})\cite{TANAKA_SURFVOLTAGEGAAS_01}. If
the density of states at this energy is large enough, the Fermi level is pinned
and hence the energy bands are bended close to the surface. Band bending is
accompanied by a surface built-in electric field and a positive charging of the
surface for p-doped semiconductors. Assuming a planar interface, the depletion
length over which the bands are shifted is $W=\sqrt{\epsilon \epsilon_0 \Phi / q
p}$, where $\epsilon=12.9$ is the dielectric constant of \ce{GaAs}, $\epsilon_0$
the vacuum permittivity and $q$ the electron charge. For
$p=$\unit{\power{10}{16}}{\centi\meter\rpcubed}, $W=$\unit{300}{\nano\meter}
($>d_{\rm QD}$). Hence the region of the PW containing the dots is fully
depleted and an electric field is present over its entire cross-section,
explaining why all studied QDs experience a drift. We have confirmed this fact
with numerical simulations. It shows that the field inside the wire is
proportional to the surface charge. Due to the conical shape of the PW a small
and almost constant $F_\|=$\unit{1.4}{\kilo\volt\per\centi\meter} component of
the electric field exists along the axis of the wire. The radial electric field
$F_\bot$ ranges from \unit{0}{\kilo\volt\per\centi\meter} on the PW axis to
\unit{-26}{\kilo\volt\per\centi\meter} at the surface. An embedded \ce{InAs} QD
subjected to this field experiences a Stark energy shift~\cite{FRY_STARKLONG_00,
GERARDOT_STARKINAS_07} $\Delta E_{\mbox{\tiny Stark}}\approx -d_\| F_\|
-\alpha_\bot F_\bot^2$, with
$d_\|\approx$\unit{40}{\micro\electronvolt\cdot\kilo\reciprocal\volt\centi\meter
} and $\alpha_\bot\approx$\unit{4}{
\micro\electronvolt\cdot\kilo\volt\rpsquared\centi\meter\squared}. We note that
a blue drift implies a reduction of the electric field, hence a diminution of
the positive net charge at the surface of the PW. 

\begin{figure}
\centering
{\includegraphics[width=7.5cm]{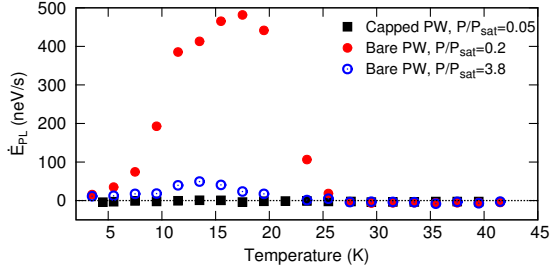}}
 \caption{(Color online) $\dot{E}_{\rm PL}$ as a function of $T$ for excitation
power $P/P_{\rm sat}=0.2$ ($\bullet$) and $P/P_{\rm sat}=3.8$ ($\odot$)
in a bare PW and for $P/P_{\rm sat}=0.05$ in a capped PW ($\blacksquare$)}
 \label{fig:temperature}
\end{figure}

We attribute this effect to the physisorption of \ce{O2} onto the surface
leading to the creation of acceptor-like surface states that capture one
electron~\cite{CHEN_OXYDATION_91, nienhaus1993physisorption,
STEVEN_ADSORPT@20K_90}. The reset of $E_{\rm PL}$ to a low value after warming
up to room temperature implies a reversible physical phenomenon and not an
irreversible chemical surface reaction. The full circles on
Fig.~\ref{fig:temperature} represent the dependence of $\dot{E}_{\rm PL}$ as a
function of $T$ between 4 and \unit{45}{\kelvin} in the same conditions as in
Fig.~\ref{fig:drift}(a). A ten-fold increase occurs from 4 to \unit{20}{\kelvin}
before the drift dramatically drops to values very close to 0 above
\unit{30}{\kelvin}. This observation  is in good agreement with previous studies
on the adsorption of \ce{O2} on \ce{GaAs}, which evidenced an adsorption peak
around \unit{20}{\kelvin}~\cite{HONIG_STICKINGO2_60,STEVEN_ADSORPT@20K_90}. The
same experiment for a capped PW  (full squares in Fig.~\ref{fig:temperature})
shows that $\dot{E}_{\rm PL}$ remains below
\unit{4}{\nano\electronvolt\per\second} over the whole temperature range,
proving the drift cancellation by the capping. The temperature dependence of
$\dot{E}_{\rm PL}$ under high excitation power ($P/P_{\rm sat}=$3.8; open
circles in Fig.~\ref{fig:temperature}) exhibits the same characteristic peak.
Its maximum value is shifted to lower temperature by $\sim$\unit{4}{\kelvin} and
is dramatically smaller than for low power excitation. The pump laser can affect
the PW in different ways: it can locally warm up the PW (explaining in
particular the peak shift at high power in Fig.~\ref{fig:temperature}),
photocreated carriers can screen the electric field
\cite{TANAKA_SURFVOLTAGEGAAS_01} or change the adsorption rate of \ce{O_2}. 

\begin{figure}
\centering
{\includegraphics[width=7.5cm]{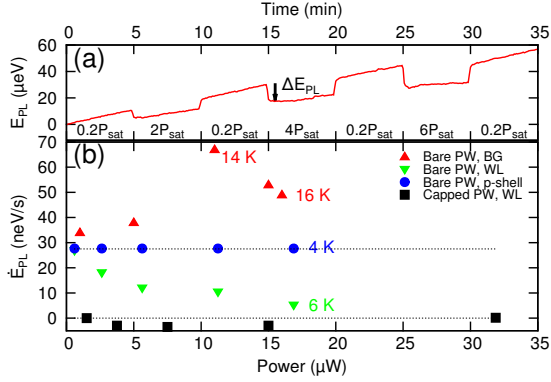}}
 \caption{(Color online)(a) $E_{\rm PL}$ of a QD versus time for various WL
excitation power (bare PW, $T=$\unit{4}{\kelvin}) (b) $\dot{E}_{\rm PL}$ versus
power of a bare PW for three excitation energies: $\rm
E_{BG}=$\unit{1.52}{\electronvolt} ($\blacktriangle$), $\rm
E_{WL}=$\unit{1.49}{\electronvolt} ($\bullet$) and $\rm
E_{pshell}=$\unit{1.42}{\electronvolt} ($\blacktriangledown$), and for a capped PW
($\blacksquare$). The PW temperature $T_{\rm PW}$ deduced from the shift $\Delta
E_{\rm PL}$ is indicated next to some points.
}
 \label{fig:powerswitch1}
\end{figure}

To better understand the phenomenon, we performed systematic power studies at
$T=$\unit{4}{\kelvin}. Figure \ref{fig:powerswitch1}(a) presents a typical
experimental run: the drift $\dot{E}_{\rm PL}$ is determined from measurement of
the emission energy over \unit{5}{\minute} at different power values. Here, we
also observe sudden shifts of the emission energy $\Delta E_{\rm PL}<0$ when the
power is increased. We attribute it to a local warming of the PW ($T_{\rm PW}\ge
T$) leading to the usual quadratic shift of the bandgap. For p-shell excitation,
the amplitude of $\Delta E_{\rm PL}$ is below the fitting uncertainty on the
peak position (\unit{0.1}{\micro\electronvolt}). In the other two cases,
$|\Delta E_{\rm PL}|$ increases linearly with power. Its magnitude is larger for
BG excitation compared to WL excitation. Similar values have been observed for
many QDs in PWs of similar size, capped or not.

The influence of power on $\dot{E}_{\rm PL}$ crucially depends on the excitation
conditions (see Fig.~\ref{fig:powerswitch1}(b)). For p-shell excitation, where
no free carrier is injected in the PW and no heating is generated, it remains
constant. In the case of WL excitation, where $T_{\rm PW}$ remains below
\unit{6}{\kelvin}, one observes a decrease of $\dot{E}_{\rm PL}$. On the other
hand for BG excitation, $T_{\rm PW}$ at high power is close to the maximum drift
temperature of Fig.~\ref{fig:temperature}. This explains why we observe an
increase of $\dot{E}_{\rm PL}$, which is dominated by temperature effect.
However it is worth noting that its maximum value is 7 times lower than the one
predicted by the low excitation results of Fig.~\ref{fig:temperature}. Hence it
is legitimate to think that, as in the case of WL excitation, the effect of
temperature is strongly inhibited by an increase of excitation power. A possible
explanation for our observations is the presence of photocreated carriers that
screen the surface built-in electric field. This phenomenon should lead to a
dramatic reduction of the electric field seen by the QD and hence induce a blue
shift of its emission energy. This is in direct contradiction with the results
of Fig.~\ref{fig:powerswitch1}(a). 
A another explanation that remains to be confirmed could be a direct influence of the carriers or excitation light
on the dynamics of adsorption and desorption of \ce{O2} or on its capture of an electron.   

The power dependence of $\dot{E}_{\rm PL}$ for capped PWs is given in
Fig.~\ref{fig:powerswitch1}(b) and is consistent with a complete suppression of
the drift. Although there may still exist a static electric field inside the PW
due to surface states at the \ce{GaAs}/\ce{Si3N4} interface, its value is frozen
and no longer depends on the adsorption of molecules on the device. Frequency
stability of the emission of a single QD is crucial to perform high precision
resonant spectroscopic studies. In our device, the QD now behaves as a stable
two-level system embedded into an optical waveguide made by the PW. It is an
ideal situation for studying non-linear effects at the single photon level and
implementing simple quantum information operation between light and a solid
state emitter\cite{GIANTNONLINEARITY_07}.

We acknowledge fruitful discussions with O. Demichel and M. Richard. This work
is supported by ANR (P3N project CAFE) and by fundation
``Nanosciences aux limites de la Nano\'electronique''. Sample
fabrication was done in the PTA and CEA LETI
MINATEC/DOPT clean rooms.


\begin{thebibliography}{0}%
\makeatletter
\providecommand \@ifxundefined [1]{%
 \@ifx{#1\undefined}
}%
\providecommand \@ifnum [1]{%
 \ifnum #1\expandafter \@firstoftwo
 \else \expandafter \@secondoftwo
 \fi
}%
\providecommand \@ifx [1]{%
 \ifx #1\expandafter \@firstoftwo
 \else \expandafter \@secondoftwo
 \fi
}%
\providecommand \natexlab [1]{#1}%
\providecommand \enquote  [1]{``#1''}%
\providecommand \bibnamefont  [1]{#1}%
\providecommand \bibfnamefont [1]{#1}%
\providecommand \citenamefont [1]{#1}%
\providecommand \href@noop [0]{\@secondoftwo}%
\providecommand \href [0]{\begingroup \@sanitize@url \@href}%
\providecommand \@href[1]{\@@startlink{#1}\@@href}%
\providecommand \@@href[1]{\endgroup#1\@@endlink}%
\providecommand \@sanitize@url [0]{\catcode `\\12\catcode `\$12\catcode
  `\&12\catcode `\#12\catcode `\^12\catcode `\_12\catcode `\%12\relax}%
\providecommand \@@startlink[1]{}%
\providecommand \@@endlink[0]{}%
\providecommand \url  [0]{\begingroup\@sanitize@url \@url }%
\providecommand \@url [1]{\endgroup\@href {#1}{\urlprefix }}%
\providecommand \urlprefix  [0]{URL }%
\providecommand \Eprint [0]{\href }%
\@ifxundefined \urlstyle {%
  \providecommand \doi  [0]{\begingroup \@sanitize@url \@doi}%
  \providecommand \@doi [1]{\endgroup \@@startlink {\doibase
  #1}doi:\discretionary {}{}{}#1\@@endlink }%
}{%
  \providecommand \doi  [0]{doi:\discretionary{}{}{}\begingroup
  \urlstyle{rm}\Url }%
}%
\providecommand \doibase [0]{http://dx.doi.org/}%
\providecommand \Doi [0]{\begingroup \@sanitize@url \@Doi }%
\providecommand \@Doi  [1]{\endgroup\@@startlink{\doibase#1}\@@Doi}%
\providecommand \@@Doi [1]{#1\@@endlink}%
\providecommand \selectlanguage [0]{\@gobble}%
\providecommand \bibinfo  [0]{\@secondoftwo}%
\providecommand \bibfield  [0]{\@secondoftwo}%
\providecommand \translation [1]{[#1]}%
\providecommand \BibitemOpen [0]{}%
\providecommand \bibitemStop [0]{}%
\providecommand \bibitemNoStop [0]{.\EOS\space}%
\providecommand \EOS [0]{\spacefactor3000\relax}%
\providecommand \BibitemShut  [1]{\csname bibitem#1\endcsname}%
\end{thebibliography}%


\begin{thebibliography}{15}%
\makeatletter
\providecommand \@ifxundefined [1]{%
 \@ifx{#1\undefined}
}%
\providecommand \@ifnum [1]{%
 \ifnum #1\expandafter \@firstoftwo
 \else \expandafter \@secondoftwo
 \fi
}%
\providecommand \@ifx [1]{%
 \ifx #1\expandafter \@firstoftwo
 \else \expandafter \@secondoftwo
 \fi
}%
\providecommand \natexlab [1]{#1}%
\providecommand \enquote  [1]{``#1''}%
\providecommand \bibnamefont  [1]{#1}%
\providecommand \bibfnamefont [1]{#1}%
\providecommand \citenamefont [1]{#1}%
\providecommand \href@noop [0]{\@secondoftwo}%
\providecommand \href [0]{\begingroup \@sanitize@url \@href}%
\providecommand \@href[1]{\@@startlink{#1}\@@href}%
\providecommand \@@href[1]{\endgroup#1\@@endlink}%
\providecommand \@sanitize@url [0]{\catcode `\\12\catcode `\$12\catcode
  `\&12\catcode `\#12\catcode `\^12\catcode `\_12\catcode `\%12\relax}%
\providecommand \@@startlink[1]{}%
\providecommand \@@endlink[0]{}%
\providecommand \url  [0]{\begingroup\@sanitize@url \@url }%
\providecommand \@url [1]{\endgroup\@href {#1}{\urlprefix }}%
\providecommand \urlprefix  [0]{URL }%
\providecommand \Eprint [0]{\href }%
\providecommand \doibase [0]{http://dx.doi.org/}%
\providecommand \selectlanguage [0]{\@gobble}%
\providecommand \bibinfo  [0]{\@secondoftwo}%
\providecommand \bibfield  [0]{\@secondoftwo}%
\providecommand \translation [1]{[#1]}%
\providecommand \BibitemOpen [0]{}%
\providecommand \bibitemStop [0]{}%
\providecommand \bibitemNoStop [0]{.\EOS\space}%
\providecommand \EOS [0]{\spacefactor3000\relax}%
\providecommand \BibitemShut  [1]{\csname bibitem#1\endcsname}%
\let\auto@bib@innerbib\@empty
\bibitem [{\citenamefont {Auff{\`e}ves-Garnier}\ \emph
  {et~al.}(2007)\citenamefont {Auff{\`e}ves-Garnier}, \citenamefont {Simon},
  \citenamefont {G\'erard},\ and\ \citenamefont
  {Poizat}}]{GIANTNONLINEARITY_07}%
  \BibitemOpen
  \bibfield  {author} {\bibinfo {author} {\bibfnamefont {A.}~\bibnamefont
  {Auff{\`e}ves-Garnier}}, \bibinfo {author} {\bibfnamefont {C.}~\bibnamefont
  {Simon}}, \bibinfo {author} {\bibfnamefont {J.-M.}\ \bibnamefont {G\'erard}},
  \ and\ \bibinfo {author} {\bibfnamefont {J.-P.}\ \bibnamefont {Poizat}},\
  }\href {\doibase 10.1103/PhysRevA.75.053823} {\bibfield  {journal} {\bibinfo
  {journal} {Phys. Rev. A}\ }\textbf {\bibinfo {volume} {75}},\ \bibinfo
  {pages} {053823} (\bibinfo {year} {2007})}\BibitemShut {NoStop}%
\bibitem [{\citenamefont {Englund}\ \emph {et~al.}(2007)\citenamefont
  {Englund}, \citenamefont {Faraon}, \citenamefont {Fushman}, \citenamefont
  {Stoltz}, \citenamefont {Petroff},\ and\ \citenamefont
  {Vuckovic}}]{EnglundVuckovic_Controllingcavityreflectivity_07}%
  \BibitemOpen
  \bibfield  {author} {\bibinfo {author} {\bibfnamefont {D.}~\bibnamefont
  {Englund}}, \bibinfo {author} {\bibfnamefont {A.}~\bibnamefont {Faraon}},
  \bibinfo {author} {\bibfnamefont {I.}~\bibnamefont {Fushman}}, \bibinfo
  {author} {\bibfnamefont {N.}~\bibnamefont {Stoltz}}, \bibinfo {author}
  {\bibfnamefont {P.}~\bibnamefont {Petroff}}, \ and\ \bibinfo {author}
  {\bibfnamefont {J.}~\bibnamefont {Vuckovic}},\ }\href
  {http://dx.doi.org/10.1038/nature06234} {\bibfield  {journal} {\bibinfo
  {journal} {Nature}\ }\textbf {\bibinfo {volume} {450}},\ \bibinfo {pages}
  {857} (\bibinfo {year} {2007})}\BibitemShut {NoStop}%
\bibitem [{\citenamefont {Friedler}\ \emph {et~al.}(2009)\citenamefont
  {Friedler}, \citenamefont {Sauvan}, \citenamefont {Hugonin}, \citenamefont
  {Lalanne}, \citenamefont {Claudon},\ and\ \citenamefont
  {G\'{e}rard}}]{FriedlerG'erard_Solid-statesinglephoton_09}%
  \BibitemOpen
  \bibfield  {author} {\bibinfo {author} {\bibfnamefont {I.}~\bibnamefont
  {Friedler}}, \bibinfo {author} {\bibfnamefont {C.}~\bibnamefont {Sauvan}},
  \bibinfo {author} {\bibfnamefont {J.~P.}\ \bibnamefont {Hugonin}}, \bibinfo
  {author} {\bibfnamefont {P.}~\bibnamefont {Lalanne}}, \bibinfo {author}
  {\bibfnamefont {J.}~\bibnamefont {Claudon}}, \ and\ \bibinfo {author}
  {\bibfnamefont {J.~M.}\ \bibnamefont {G\'{e}rard}},\ }\href {\doibase
  10.1364/OE.17.002095} {\bibfield  {journal} {\bibinfo  {journal} {Opt.
  Express}\ }\textbf {\bibinfo {volume} {17}},\ \bibinfo {pages} {2095}
  (\bibinfo {year} {2009})}\BibitemShut {NoStop}%
\bibitem [{\citenamefont {Claudon}\ \emph {et~al.}(2010)\citenamefont
  {Claudon}, \citenamefont {Bleuse}, \citenamefont {Malik}, \citenamefont
  {Bazin}, \citenamefont {Jaffrennou}, \citenamefont {Gregersen}, \citenamefont
  {Sauvan}, \citenamefont {Lalanne},\ and\ \citenamefont
  {G{\'e}rard}}]{ClaudonGerard_highlyefficientsingle-photon_10}%
  \BibitemOpen
  \bibfield  {author} {\bibinfo {author} {\bibfnamefont {J.}~\bibnamefont
  {Claudon}}, \bibinfo {author} {\bibfnamefont {J.}~\bibnamefont {Bleuse}},
  \bibinfo {author} {\bibfnamefont {N.~S.}\ \bibnamefont {Malik}}, \bibinfo
  {author} {\bibfnamefont {M.}~\bibnamefont {Bazin}}, \bibinfo {author}
  {\bibfnamefont {P.}~\bibnamefont {Jaffrennou}}, \bibinfo {author}
  {\bibfnamefont {N.}~\bibnamefont {Gregersen}}, \bibinfo {author}
  {\bibfnamefont {C.}~\bibnamefont {Sauvan}}, \bibinfo {author} {\bibfnamefont
  {P.}~\bibnamefont {Lalanne}}, \ and\ \bibinfo {author} {\bibfnamefont
  {J.-M.}\ \bibnamefont {G{\'e}rard}},\ }\href
  {http://dx.doi.org/10.1038/nphoton.2009.287} {\bibfield  {journal} {\bibinfo
  {journal} {Nat Photon}\ }\textbf {\bibinfo {volume} {4}},\ \bibinfo {pages}
  {174} (\bibinfo {year} {2010})}\BibitemShut {NoStop}%
\bibitem [{\citenamefont {Babinec}\ \emph {et~al.}(2010)\citenamefont
  {Babinec}, \citenamefont {M.}, \citenamefont {Khan}, \citenamefont {Zhang},
  \citenamefont {Maze}, \citenamefont {Hemmer},\ and\ \citenamefont
  {Loncar}}]{BabinecLoncar_diamondnanowiresingle-photon_10}%
  \BibitemOpen
  \bibfield  {author} {\bibinfo {author} {\bibfnamefont {T.~M.}\ \bibnamefont
  {Babinec}}, \bibinfo {author} {\bibfnamefont {H.~J.}\ \bibnamefont {M.}},
  \bibinfo {author} {\bibfnamefont {M.}~\bibnamefont {Khan}}, \bibinfo {author}
  {\bibfnamefont {Y.}~\bibnamefont {Zhang}}, \bibinfo {author} {\bibfnamefont
  {J.~R.}\ \bibnamefont {Maze}}, \bibinfo {author} {\bibfnamefont {P.~R.}\
  \bibnamefont {Hemmer}}, \ and\ \bibinfo {author} {\bibfnamefont
  {M.}~\bibnamefont {Loncar}},\ }\href {http://dx.doi.org/10.1038/nnano.2010.6}
  {\bibfield  {journal} {\bibinfo  {journal} {Nat Nano}\ }\textbf {\bibinfo
  {volume} {5}},\ \bibinfo {pages} {195} (\bibinfo {year} {2010})}\BibitemShut
  {NoStop}%
\bibitem [{\citenamefont {Bleuse}\ \emph {et~al.}(2011)\citenamefont {Bleuse},
  \citenamefont {Claudon}, \citenamefont {Creasey}, \citenamefont {Malik},
  \citenamefont {G\'erard}, \citenamefont {Maksymov}, \citenamefont {Hugonin},\
  and\ \citenamefont {Lalanne}}]{NANOWIRESECONTROLBLEUSE11}%
  \BibitemOpen
  \bibfield  {author} {\bibinfo {author} {\bibfnamefont {J.}~\bibnamefont
  {Bleuse}}, \bibinfo {author} {\bibfnamefont {J.}~\bibnamefont {Claudon}},
  \bibinfo {author} {\bibfnamefont {M.}~\bibnamefont {Creasey}}, \bibinfo
  {author} {\bibfnamefont {N.~S.}\ \bibnamefont {Malik}}, \bibinfo {author}
  {\bibfnamefont {J.-M.}\ \bibnamefont {G\'erard}}, \bibinfo {author}
  {\bibfnamefont {I.}~\bibnamefont {Maksymov}}, \bibinfo {author}
  {\bibfnamefont {J.-P.}\ \bibnamefont {Hugonin}}, \ and\ \bibinfo {author}
  {\bibfnamefont {P.}~\bibnamefont {Lalanne}},\ }\href {\doibase
  10.1103/PhysRevLett.106.103601} {\bibfield  {journal} {\bibinfo  {journal}
  {Phys. Rev. Lett.}\ }\textbf {\bibinfo {volume} {106}},\ \bibinfo {pages}
  {103601} (\bibinfo {year} {2011})}\BibitemShut {NoStop}%
\bibitem [{\citenamefont {Demichel}\ \emph {et~al.}(2010)\citenamefont
  {Demichel}, \citenamefont {Heiss}, \citenamefont {Bleuse}, \citenamefont
  {Mariette},\ and\ \citenamefont {i~Morral}}]{DEMICHEL_SURFACEGAAS_10}%
  \BibitemOpen
  \bibfield  {author} {\bibinfo {author} {\bibfnamefont {O.}~\bibnamefont
  {Demichel}}, \bibinfo {author} {\bibfnamefont {M.}~\bibnamefont {Heiss}},
  \bibinfo {author} {\bibfnamefont {J.}~\bibnamefont {Bleuse}}, \bibinfo
  {author} {\bibfnamefont {H.}~\bibnamefont {Mariette}}, \ and\ \bibinfo
  {author} {\bibfnamefont {A.~F.}\ \bibnamefont {i~Morral}},\ }\href {\doibase
  10.1063/1.3519980} {\bibfield  {journal} {\bibinfo  {journal} {Appl. Phys.
  Lett.}\ }\textbf {\bibinfo {volume} {97}},\ \bibinfo {eid} {201907} (\bibinfo
  {year} {2010})}\BibitemShut {NoStop}%
\bibitem [{\citenamefont {Dan}\ \emph {et~al.}(2011)\citenamefont {Dan},
  \citenamefont {Seo}, \citenamefont {Takei}, \citenamefont {Meza},
  \citenamefont {Javey},\ and\ \citenamefont {Crozier}}]{SURFACEEFFECTDAN11}%
  \BibitemOpen
  \bibfield  {author} {\bibinfo {author} {\bibfnamefont {Y.}~\bibnamefont
  {Dan}}, \bibinfo {author} {\bibfnamefont {K.}~\bibnamefont {Seo}}, \bibinfo
  {author} {\bibfnamefont {K.}~\bibnamefont {Takei}}, \bibinfo {author}
  {\bibfnamefont {J.~H.}\ \bibnamefont {Meza}}, \bibinfo {author}
  {\bibfnamefont {A.}~\bibnamefont {Javey}}, \ and\ \bibinfo {author}
  {\bibfnamefont {K.~B.}\ \bibnamefont {Crozier}},\ }\href {\doibase
  10.1021/nl201179n} {\bibfield  {journal} {\bibinfo  {journal} {Nano Letters}\
  }\textbf {\bibinfo {volume} {11}},\ \bibinfo {pages} {2527} (\bibinfo {year}
  {2011})}\BibitemShut {NoStop}%
\bibitem [{\citenamefont {Tanaka}\ \emph {et~al.}(2001)\citenamefont {Tanaka},
  \citenamefont {More}, \citenamefont {Murakami}, \citenamefont {Itoh},
  \citenamefont {Fujii},\ and\ \citenamefont
  {Kamada}}]{TANAKA_SURFVOLTAGEGAAS_01}%
  \BibitemOpen
  \bibfield  {author} {\bibinfo {author} {\bibfnamefont {S.}~\bibnamefont
  {Tanaka}}, \bibinfo {author} {\bibfnamefont {S.~D.}\ \bibnamefont {More}},
  \bibinfo {author} {\bibfnamefont {J.}~\bibnamefont {Murakami}}, \bibinfo
  {author} {\bibfnamefont {M.}~\bibnamefont {Itoh}}, \bibinfo {author}
  {\bibfnamefont {Y.}~\bibnamefont {Fujii}}, \ and\ \bibinfo {author}
  {\bibfnamefont {M.}~\bibnamefont {Kamada}},\ }\href {\doibase
  10.1103/PhysRevB.64.155308} {\bibfield  {journal} {\bibinfo  {journal} {Phys.
  Rev. B}\ }\textbf {\bibinfo {volume} {64}},\ \bibinfo {pages} {155308}
  (\bibinfo {year} {2001})}\BibitemShut {NoStop}%
\bibitem [{\citenamefont {Fry}\ \emph {et~al.}(2000)\citenamefont {Fry},
  \citenamefont {Itskevich}, \citenamefont {Mowbray}, \citenamefont {Skolnick},
  \citenamefont {Finley}, \citenamefont {Barker}, \citenamefont {O'Reilly},
  \citenamefont {Wilson}, \citenamefont {Larkin}, \citenamefont {Maksym},
  \citenamefont {Hopkinson}, \citenamefont {Al-Khafaji}, \citenamefont {David},
  \citenamefont {Cullis}, \citenamefont {Hill},\ and\ \citenamefont
  {Clark}}]{FRY_STARKLONG_00}%
  \BibitemOpen
  \bibfield  {author} {\bibinfo {author} {\bibfnamefont {P.~W.}\ \bibnamefont
  {Fry}}, \bibinfo {author} {\bibfnamefont {I.~E.}\ \bibnamefont {Itskevich}},
  \bibinfo {author} {\bibfnamefont {D.~J.}\ \bibnamefont {Mowbray}}, \bibinfo
  {author} {\bibfnamefont {M.~S.}\ \bibnamefont {Skolnick}}, \bibinfo {author}
  {\bibfnamefont {J.~J.}\ \bibnamefont {Finley}}, \bibinfo {author}
  {\bibfnamefont {J.~A.}\ \bibnamefont {Barker}}, \bibinfo {author}
  {\bibfnamefont {E.~P.}\ \bibnamefont {O'Reilly}}, \bibinfo {author}
  {\bibfnamefont {L.~R.}\ \bibnamefont {Wilson}}, \bibinfo {author}
  {\bibfnamefont {I.~A.}\ \bibnamefont {Larkin}}, \bibinfo {author}
  {\bibfnamefont {P.~A.}\ \bibnamefont {Maksym}}, \bibinfo {author}
  {\bibfnamefont {M.}~\bibnamefont {Hopkinson}}, \bibinfo {author}
  {\bibfnamefont {M.}~\bibnamefont {Al-Khafaji}}, \bibinfo {author}
  {\bibfnamefont {J.~P.~R.}\ \bibnamefont {David}}, \bibinfo {author}
  {\bibfnamefont {A.~G.}\ \bibnamefont {Cullis}}, \bibinfo {author}
  {\bibfnamefont {G.}~\bibnamefont {Hill}}, \ and\ \bibinfo {author}
  {\bibfnamefont {J.~C.}\ \bibnamefont {Clark}},\ }\href {\doibase
  10.1103/PhysRevLett.84.733} {\bibfield  {journal} {\bibinfo  {journal} {Phys.
  Rev. Lett.}\ }\textbf {\bibinfo {volume} {84}},\ \bibinfo {pages} {733}
  (\bibinfo {year} {2000})}\BibitemShut {NoStop}%
\bibitem [{\citenamefont {Gerardot}\ \emph {et~al.}(2007)\citenamefont
  {Gerardot}, \citenamefont {Seidl}, \citenamefont {Dalgarno}, \citenamefont
  {Warburton}, \citenamefont {Granados}, \citenamefont {Garcia}, \citenamefont
  {Kowalik}, \citenamefont {Krebs}, \citenamefont {Karrai}, \citenamefont
  {Badolato},\ and\ \citenamefont {Petroff}}]{GERARDOT_STARKINAS_07}%
  \BibitemOpen
  \bibfield  {author} {\bibinfo {author} {\bibfnamefont {B.~D.}\ \bibnamefont
  {Gerardot}}, \bibinfo {author} {\bibfnamefont {S.}~\bibnamefont {Seidl}},
  \bibinfo {author} {\bibfnamefont {P.~A.}\ \bibnamefont {Dalgarno}}, \bibinfo
  {author} {\bibfnamefont {R.~J.}\ \bibnamefont {Warburton}}, \bibinfo {author}
  {\bibfnamefont {D.}~\bibnamefont {Granados}}, \bibinfo {author}
  {\bibfnamefont {J.~M.}\ \bibnamefont {Garcia}}, \bibinfo {author}
  {\bibfnamefont {K.}~\bibnamefont {Kowalik}}, \bibinfo {author} {\bibfnamefont
  {O.}~\bibnamefont {Krebs}}, \bibinfo {author} {\bibfnamefont
  {K.}~\bibnamefont {Karrai}}, \bibinfo {author} {\bibfnamefont
  {A.}~\bibnamefont {Badolato}}, \ and\ \bibinfo {author} {\bibfnamefont
  {P.~M.}\ \bibnamefont {Petroff}},\ }\href {\doibase 10.1063/1.2431758}
  {\bibfield  {journal} {\bibinfo  {journal} {Appl. Phys. Lett.}\ }\textbf
  {\bibinfo {volume} {90}},\ \bibinfo {eid} {041101} (\bibinfo {year}
  {2007})}\BibitemShut {NoStop}%
\bibitem [{\citenamefont {Chen}\ \emph {et~al.}(1991)\citenamefont {Chen},
  \citenamefont {Stepniak}, \citenamefont {Seo}, \citenamefont {Harvey},\ and\
  \citenamefont {Weaver}}]{CHEN_OXYDATION_91}%
  \BibitemOpen
  \bibfield  {author} {\bibinfo {author} {\bibfnamefont {Y.}~\bibnamefont
  {Chen}}, \bibinfo {author} {\bibfnamefont {F.}~\bibnamefont {Stepniak}},
  \bibinfo {author} {\bibfnamefont {J.~M.}\ \bibnamefont {Seo}}, \bibinfo
  {author} {\bibfnamefont {S.~E.}\ \bibnamefont {Harvey}}, \ and\ \bibinfo
  {author} {\bibfnamefont {J.~H.}\ \bibnamefont {Weaver}},\ }\href {\doibase
  10.1103/PhysRevB.43.12086} {\bibfield  {journal} {\bibinfo  {journal} {Phys.
  Rev. B}\ }\textbf {\bibinfo {volume} {43}},\ \bibinfo {pages} {12086}
  (\bibinfo {year} {1991})}\BibitemShut {NoStop}%
\bibitem [{\citenamefont {Nienhaus}\ and\ \citenamefont
  {M{\"o}nch}(1993)}]{nienhaus1993physisorption}%
  \BibitemOpen
  \bibfield  {author} {\bibinfo {author} {\bibfnamefont {H.}~\bibnamefont
  {Nienhaus}}\ and\ \bibinfo {author} {\bibfnamefont {W.}~\bibnamefont
  {M{\"o}nch}},\ }\href@noop {} {\bibfield  {journal} {\bibinfo  {journal}
  {Appl. Surf. Sci.}\ }\textbf {\bibinfo {volume} {65}},\ \bibinfo {pages}
  {632} (\bibinfo {year} {1993})}\BibitemShut {NoStop}%
\bibitem [{\citenamefont {Anderson}\ \emph {et~al.}(1990)\citenamefont
  {Anderson}, \citenamefont {Komeda}, \citenamefont {Seo}, \citenamefont
  {Capasso}, \citenamefont {Waddill}, \citenamefont {Benning},\ and\
  \citenamefont {Weaver}}]{STEVEN_ADSORPT@20K_90}%
  \BibitemOpen
  \bibfield  {author} {\bibinfo {author} {\bibfnamefont {S.~G.}\ \bibnamefont
  {Anderson}}, \bibinfo {author} {\bibfnamefont {T.}~\bibnamefont {Komeda}},
  \bibinfo {author} {\bibfnamefont {J.~M.}\ \bibnamefont {Seo}}, \bibinfo
  {author} {\bibfnamefont {C.}~\bibnamefont {Capasso}}, \bibinfo {author}
  {\bibfnamefont {G.~D.}\ \bibnamefont {Waddill}}, \bibinfo {author}
  {\bibfnamefont {P.~J.}\ \bibnamefont {Benning}}, \ and\ \bibinfo {author}
  {\bibfnamefont {J.~H.}\ \bibnamefont {Weaver}},\ }\href {\doibase
  10.1103/PhysRevB.42.5082} {\bibfield  {journal} {\bibinfo  {journal} {Phys.
  Rev. B}\ }\textbf {\bibinfo {volume} {42}},\ \bibinfo {pages} {5082}
  (\bibinfo {year} {1990})}\BibitemShut {NoStop}%
\bibitem [{\citenamefont {Honig}\ and\ \citenamefont
  {Hook}(1960)}]{HONIG_STICKINGO2_60}%
  \BibitemOpen
  \bibfield  {author} {\bibinfo {author} {\bibfnamefont {R.~E.}\ \bibnamefont
  {Honig}}\ and\ \bibinfo {author} {\bibfnamefont {R.~O.}\ \bibnamefont
  {Hook}},\ }\href@noop {} {\bibfield  {journal} {\bibinfo  {journal} {RCA
  Rev.}\ }\textbf {\bibinfo {volume} {21}},\ \bibinfo {pages} {360} (\bibinfo
  {year} {1960})}\BibitemShut {NoStop}%
\end{thebibliography}
%

\end{document}